\documentclass[12pt]{article}
\title{Implementing an Observatory Control System-I. A Generic approach}
\author{Sunu Engineer\\Inter University Centre for Astronomy and Astrophysics\\Pune, India \thanks{sunu@iucaa.ernet.in}}
\date{\today}

\begin{document}
\maketitle
\abstract{An architectural framework for implementing a distributed observatory control system is presented here. It has been partially realized and tested in the 2m optical and infrared  Observatory at Pune, India.}
\section{Introduction}
An observatory, irrespective of wavelengths of observations, is a complex instrument consisting of a large number of interacting parts. The creation of a such a structure is a difficult problem, fraught with issues of various kinds. However, there is a great deal in common between the observatories of past, present and possibly future and it is important to create a framework which takes advantage of this commonality. In spite of the fact that scientists as a rule tend towards creating unified standardized approaches to a subject, such a standard is seemingly lacking in the astronomy related engineering and instrumentation field or is not very popular. In this paper we attempt to address the issue of consolidation of many years of experience in building such complex instruments, and creation of a framework which addresses the practical needs of the community, spanning the gamut of observatories from earth based to space based ones.
\section{Architecture of an Observatory}
The functional requirements of an observatory determines to a large extent both the physical and software architecture of the system. Unlike older observatories which had many instruments primarily interconnected via the operators, modern observatories have an internal structure which consists of parts interlinked through media capable of transmitting digital signals. The required analog conversions are done at possibly the last step, thus allowing the observatory to be modeled as a large asynchronous digital network.
\par
Given this the problem that we are trying to address is the following: Is it possible to define a generic set of components and interconnects, starting from hardware to users of various kinds, which will allow the observatory to be modeled and built with a minimum of effort and a maximum of stable, reliable functionality. It is also desired that the structures be generic enough to permit an easy mix and match between the different observatories, while maintaining the ability to be customized to the requirements of a particular entity.
\par
In order to address this, it is proposed that the observatory be modeled and described as a set of interacting objects (in the computer programming sense of the word) and such a model be implemented at all levels by creating hardware and software components which can be put in one to one correspondence with the model. In this paper we present one such model. 
\par
The model is intended to be an evolutionary structure, permitting instantiation of it with variations at various sites and a feedback and evolve methodology defined within.
\section{The Common Object Framework}
In order to facilitate such an architecture one needs to adopt a standard framework for designing and implementing the Observatory Control System. Moreover the standardisation attempt should extend itself to hardware (the common parts of it at least) as well as the test and calibration framework.
\par
What is proposed here in the current paper is such a framework for a distributed observatory system. In forthcoming papers we detail the implementation aspects as applied in the IUCAA observatory project. 
\par
A modern observatory consists of interconnected digital systems from motion control to data acquisition and archiving, irrespective of wavelength or location of the system. Going further many of them are designed to operate with minimal or no manual intervention, such in the case of space observatories or robotic ones. Building and deploying such a complex system to reach level of high precision and noise free observations is currently an expensive and time consuming task.  However there have been a large number of them of all sizes and categories built and deployed on Earth as well as in nearby orbits in space. The knowledge and experience that has accumulated from such successful and unsuccessful endeavours is enormous. However the knowledge has not been translated into generic patterns \cite{BigFour} which can be used to create practical and stable platforms. There is a requirement for a centralised repository of auxiliary information such as the issues faced during development and deployment as well as data regarding long term use of the instruments and so on.In the software engineering field many approaches have been developed to create patterns for distributed and concurrent systems to which class observatories certainly belong. \cite{concurrent}.  In a like manner we try to describe a methodology for practically constructing these systems. It has been suggested that since the technology is constantly changing and it is difficult for the astronomy community to concentrate the resources from different domains of expertise, it would be efficient to have such models implemented using currently established software development companies. It would also allow experienced resource management frameworks to be introduced into the picture for timely and reliable creation. (Prof. Ajit Kembhavi, private communication)
\par
There are many aspects to be considered in the common framework. Other than achieving the desired functionality, the system must also be open to probing and diagnostics as well as maintenance and other management issues. Issues related to calibration and performance analysis of the system are also critical to successful deployment. 
\par
\subsection{The basic model}
The observatory is designed by assembling together various components (almost like a Lego kit). A component is defined herein as an entity consisting of state and behaviour, implemented both in hardware and software. In some senses it is a unification of hardware and software under the object oriented paradigm. The component also incorporates along with other functional aspects related to application, an interconnect architecture, including the hardware and the software protocol stack. Thus it is akin to the transputer in the world of microprocessors.  In practice, the notion of ``component'' is defined not at the level of the controller that controls the hardware, but more at the level of the PC that interfaces to the hardware device on one side and meshes into the observatory control system via the transport layer interface.  Thus an important aspect of the state of the ``component'' is the interconnect address. It is proposed that this communication layer be TCP--UDP/IP, the standard protocol stack of the Internet. 
\par
Over this communication link, a message based, event driven architecture is defined which allows the whole observatory to be built up by integrating these components. Self descriptive messages in XML or in alternative formats can be sent across these channels in point to point, broadcast or to a central message router and so on such that the sender and recipient can effect either synchronous or asynchronous changes of state based on them. A suite of messages relevant for this model in XML format is presented later. 
\par
Messages can have different degrees of genericity and so can the family of handlers which process the messages. There is a hierarchy of handlers starting from the most general to the most specific as in standard protocol stacks. 
\par
\subsection{Deployment and Initialization}
We begin by assuming that the final entity built out of these components have been deployed in a required spatial pattern. However we must address the issue of initialization of the same.  The proposed methodology is to use a central configuration server to serve the configuration information for each component via the message protocol and update itself on receiving initialization confirmation from the components. The server  is based on LDAP although any other directory/database server could be utilized for the same purpose. However the schema that is used to store and communicate the configuration information, in the form of a descriptor must be standardised. The initialization phase is  modeled akin to  the discovery phase of a protocol such as USB. 
\par
Initialization also involves creation of global data which can be used for communication as well as synchronization by the components.  A service providing this facility for a distributed system is a component of the architecture like others.  
\par
However there is an issue of time ordering of the initialization. This is an important consideration which is addressed in the implementation stage, where the server masterminds the order in which the components are initialized inspite of the arbitrary ordering of startup in physical time.
\subsection{The component architecture}
As is clear from the above discussion, there are two degrees of freedom in design and implementation. By evolving a standard for both we should limit the variability in implementation. 
\par
We begin by designing an observatory profile which defines and characterises the components that constitute an observatory. We further define and categorise the messages that can be sent and received by the components, there by producing a constrained environment which meets our needs.  The system is designed and constructed to be an open system allowing  the addition of more ratified messages and components as well as leaving some free space in the object definitions and message definitions. The algorithms that the message handlers implement are also defined to a great degree by the standard but more in line with directives, rather than mandates.
\section{The Observatory Profile}
The observatory profile consists of the list of components and the messages that they transmit and receive. The profile defines the basic structure of the component and delineates an object framework for creating and classifying them. It also defines the functionality in detail along with the algorithms that are used/recommended to implement the functionality.  In this section we will briefly discuss the observatory profile which is detailed in a separate paper.
The profile consists of a set of other profiles which individually detail the objects and their behavioural and structural constraints. 
\begin{enumerate}
\item{Telescope System Profile: A detailed set of objects that encapsulate the telescope system and its behaviour including the autoguider and pointing models. }
\item{Enclosure System Profile: Set of objects that control the enclosure of the telescope system and individual instruments.}
\item{Instrument  Profile: Set of instruments such as CCD cameras and other Imagers, Spectroscopes, Polarimeters and others.}
\item{Data archiving and Processing Profile: Objects to archive data in raw and processed forms (from image processing pipelines deployed on clusters and grids) such that query, retrieval etc. is uniform, distributed and simple.}
\item{Management System Profile: The large set of objects which are used to manage the observatory and its components such as loggers, monitors and management stations by means of which the operators can interact with the observatory system.}
\item{Configuration and security repository Profile: The central repository storing configuration information regarding the components of the system, the users and their authorizations and so on.}
\item{External Environment and Meteorology System Profile: The set of objects comprising the functionality required to access, control and monitor the external conditions such as electrical and mechanical power, weather conditions, infrastructure (networks for instance) conditions and so on. }
\end{enumerate}
Each individual profile is discussed in greater detail in later papers. However in the following sections we will briefly discuss some non standard subsystems. 
\section{The Management and Configuration subsystems}
In order to keep such a complex system running smoothly as well as make it extensible and maintainable one must have security and management capabilities defined in the the infrastructure itself. A management model based on Common Infrastructure for Management (CIM) \cite{cim} over HTTP may be utilised to ensure that a stable generic management platform is integrated into the design. 
\par
In a like manner a security system suitable for such a distributed architecture must be implemented and the security credentials must be movable so that distributed resources can be accessed transparently. These issues will be discussed in detail in the later work. 
\section{The data archiving and processing system}
Since the {\it raison d'etre} of the observatory is to produce data for science the data archiving and processing subsystems are extremely important. There are many different archiving structures and formats and a few established tools and patterns for astronomical data processing.  One of the primary products of this track which has been evolving over time is the virtual observatory project.\cite{vobs} 
\par
The data archiving component is designed to be conformant with the ongoing efforts in this field and will be detailed in a later paper. The  retrieval, processing and caching of data required for creating distributed, streaming mode architectures suitable for parallel or grid based data processing and integrating the same into existing applications such as IRAF and MIDAS will also be discussed later.  
\section{Conclusions}
In this paper we present the principles and architecture of a distributed observatory control system in a generic form. We propose that such a standardised componentised architecture be used to simulate, prototype and build the observatories in all wavelengths. We propose a language independent implementation of the objects in the object oriented framework which describes both the components as well as the gluing structures as objects. In succeeding papers we discuss the actual implementation in greater detail.
\section{Acknowledgements}
The author wishes to thank IUCAA and his colleagues in the instrumentation division for much help in the implementation of this architecture. Many issues and problems were understood and resolved during discussions with Prof. S.N. Tandon of IUCAA.  We also benefited greatly from extended discussions on many aspects with Prof. Padmanabhan of IUCAA and we thank him for all the guidance.

\end{document}